\documentclass[12pt,reqno]{article}
\usepackage{amsmath}

\allowdisplaybreaks
\numberwithin{equation}{section}

\renewcommand{\d}{\mathrm{d}}
\newcommand{\I}{\mathrm{i}}
\newcommand{\e}{\mathrm{e}}
\newcommand{\A}{{\hat{A}}}
\newcommand{\B}{{\hat{B}}}
\newcommand{\C}{{\hat{C}}}
\newcommand{\D}{{\hat{D}}}
\newcommand{\cD}{\mathcal{D}}
\newcommand{\cE}{\mathcal{E}}
\newcommand{\cF}{\mathcal{F}}
\newcommand{\cG}{\mathcal{G}}
\newcommand{\cH}{\mathcal{H}}
\newcommand{\cL}{\mathcal{L}}
\newcommand{\cR}{\mathcal{R}}
\newcommand{\ab}{{\bar{a}}}
\newcommand{\bb}{{\bar{b}}}
\newcommand{\cb}{{\bar{c}}}

\newcommand{\ga}{\gamma}
\newcommand{\de}{\delta}
\newcommand{\ep}{\varepsilon}
\newcommand{\eps}{\epsilon}
\newcommand{\beps}{\bar{\eps}}
\newcommand{\si}{\sigma}
\newcommand{\bsi}{\bar{\sigma}}
\newcommand{\la}{\lambda}
\newcommand{\bla}{\bar{\lambda}}
\newcommand{\2}{\sqrt{2\,}}
\newcommand{\p}{\partial}
\newcommand{\bpsi}{\bar{\psi}}
\newcommand{\tab}{\quad\,}
\newcommand{\half}{\tfrac{1}{2}}
\newcommand{\ihalf}{\tfrac{\mathrm{i}}{2}}
\newcommand{\quart}{\tfrac{1}{4}}
\newcommand{\V}[2]{V_{\!#1}^{#2}}
\newcommand{\W}[2]{W_{\!#1}^{#2}}
\newcommand{\bV}[2]{\bar{V}_{\!#1}^{#2}}
\newcommand{\bW}[2]{\bar{W}_{\!#1}^{#2}}
\newcommand{\G}[3]{\Gamma_{\!#1}{}^{#2}{}_{#3}}
\newcommand{\frc}[2]{\frac{\raisebox{-2pt}{$#1$}}{#2}}
\newcommand{\tbs}[1]{\overset{\leftrightarrow}{#1}}

\begin{document} 

\begin{flushright} \small
 ITP--UU--03/06 \\ SPIN--03/04 \\ TUW--03--07 \\ hep-th/0303048
\end{flushright}
\bigskip

\begin{center}
 {\large\bfseries N=2 Supersymmetric Scalar-Tensor Couplings} \\[5mm]
 Ulrich Theis$^1$ and Stefan Vandoren$^2$ \\[3mm]
 {\small\slshape
 $^1$Institute for Theoretical Physics, Vienna University of
 Technology, \\ Wiedner Hauptstrasse 8--10, A-1040 Vienna, Austria \\
 {\upshape\ttfamily theis@hep.itp.tuwien.ac.at}\\[3mm]
 $^2$Institute for Theoretical Physics \emph{and} Spinoza Instituut \\
 Utrecht University, 3508 TD Utrecht, The Netherlands \\
 {\upshape\ttfamily S.Vandoren@phys.uu.nl}}
\end{center}
\vspace{5mm}

\hrule\bigskip

\centerline{\bfseries Abstract} \medskip

We determine the general coupling of a system of scalars and
antisymmetric tensors, with at most two derivatives and undeformed gauge
transformations, for both rigid and local $N=2$ supersymmetry in
four-dimensional spacetime. Our results cover interactions of hyper,
tensor and double-tensor multiplets and apply among others to Calabi-Yau
threefold compactifications of Type~II supergravities. As an example, we
give the complete Lagrangian and supersymmetry transformation rules of
the double-tensor multiplet dual to the universal hypermultiplet.

\bigskip\hrule\bigskip

\section{Introduction}

It is well-known that gauge fields of form degree $D-2$ (antisymmetric
tensors) in $D$ dimensions represent one bosonic degree of freedom, and
that often this degree of freedom can alternatively be described in
terms of a scalar field. The duality transformation relating the two
formulations of the same physics exchanges Bianchi identities and field
equations. Such tensors arise for instance in compactifications of
superstring theories to four dimensions as descendants of the
Neveu-Schwarz two-form and the Ramond-Ramond $p$-forms. Usually one
prefers to dualize these two-forms in four dimensions into scalars, in
order to understand the full duality group of the theory and to exhibit
the geometric structure of the resulting moduli spaces. In some cases,
however, it is convenient, or even necessary, to keep the tensors as
they naturally appear in string theory. One example is when there are
bare gauge potentials, not appearing only through their field strength.
This complicates, or might even obstruct, the dualization procedure.
Another example is the Euclidean formulation of string theory and its
compactifications, as used in instanton calculations: The Euclidean
scalar formulation typically suffers from an indefinite action which
invalidates the semiclassical approximation, whereas the dual tensor
formulation has a positive semi-definite action. Such a situation occurs
for instance for the ten-dimensional D-instanton in Type~IIB string
theory \cite{GGP}, but also for instantons that contribute to the
universal hypermultiplet effective action in four dimensions \cite{TV}.
Therefore, it is necessary to have a better understanding of general
scalar-tensor couplings, and the aim of this paper is to construct their
Lagrangians and transformation rules with $N=2$ supersymmetry in four
spacetime dimensions.

This paper was motivated by Calabi-Yau threefold compactifications of
Type~II string theories, which yield matter-coupled $N=2$ supergravity
as the low energy effective theory. In the absence of internal fluxes,
the vector multiplets arising from the compactification do not couple to
the other matter fields and can consistently be truncated. For Type~IIA,
there remain $h_{1,2}$ hypermultiplets and one tensor multiplet
(containing three scalars and one tensor). For Type~IIB, one has
$h_{1,1}$ tensor multiplets and one double-tensor multiplet (containing
two scalars and two tensors). The tensor multiplet in IIA and the
double-tensor multiplet in IIB are universal, in that they appear for
all choices of the Calabi-Yau manifold and do not depend on its moduli.
The generic situation is that one gets a complicated four-dimensional
low energy supergravity action with interacting scalars and tensors.
Superspace effective actions for Type~II strings were derived in
\cite{BS}, and we are interested in the on-shell component formulation
that include also the double-tensor multiplet. At tree level, the
bosonic terms of Type~IIA and Type~IIB supergravity on a Calabi-Yau were
determined in \cite{BCF} and \cite{BGHL}, respectively, but the
fermionic terms and the supersymmetry transformations are still lacking.
While the general coupling of an arbitrary number of hypermultiplets to
$N=2$ supergravity was determined in \cite{BW}, for tensor or
double-tensor multiplets a similar program has not been carried out so
far. To our knowledge, the latter has in fact never been fully coupled
to supergravity before, while at least locally supersymmetric versions
of tensor multiplets (in the superconformal approach) have appeared in
the literature in \cite{dW,BS,dWRV}. In this article we close these
gaps. We do this by employing the duality to hypermultiplets: The
general $N=2$ supersymmetric system of $n_T$ tensors and a
number\footnote{This number is not arbitrary but restricted by
supersymmetry to be a multiple of four minus $n_T$.} of scalars, where
the former are subject to standard field-independent gauge
transformations, is equivalent to a set of hypermultiplets. Their
scalars will parametrize a target space whose metric has $n_T$ commuting
isometries (in appropriate coordinates these are just the Peccei-Quinn
shift symmetries). Hence, the general $N=2$ supersymmetric coupling of
tensors to scalars can be derived by dualizing the general system of
interacting hypermultiplets with a number of abelian isometries.

Our insistence on undeformed gauge transformations ensures that the
tensors occur only through their field strengths\footnote{There are no
Chern-Simons terms for 2-forms in four dimensions.}. At least for rigid
supersymmetry, this does not yield the most general self-interactions of
tensors. There are in addition Freedman-Townsend models that involve
bare gauge potentials\footnote{To first order in the coupling constant,
the interactions are of the form $gf^{IJK}B_I*\!\d B_J*\!\d B_K$, where
$f^{IJK}$ are the structure constants of some Lie algebra.}, but that
are still dual to nonlinear sigma models \cite{OP}. The distinguishing
feature of Freedman-Townsend couplings is that the gauge transformations
of the tensors are deformed and become field-dependent. As was argued in
\cite{B1}, there seem to exist globally supersymmetric self-interactions
of the double-tensor multiplet that are of this form. We exclude such
models from our consideration, but this does not significantly reduce
the applicability of our results, since we are mainly interested in
Calabi-Yau compactifications of Type~II strings, where such
Freedman-Townsend couplings do not appear.

The paper is organized as follows: In section \ref{sect_rigid} we first
give a brief review of globally supersymmetric hypermultiplet
interactions and discuss commuting isometries necessary for the
dualization of scalars into tensors. We then explain in detail the
dualization procedure via the gauging of isometries and the derivation
of the supersymmetry transformations for the dual theory. The reader not
interested in this derivation can go directly to the last subsection,
where the scalar-tensor multiplet couplings are directly given, without
referring to the dual hypermultiplet system. In section \ref{sect_local}
we repeat these steps (in less detail) for local supersymmetry, and in
section \ref{sect_DTM} we discuss the example of the universal
hypermultiplet and its dual formulation in terms of the double-tensor
multiplet.

\section{Rigid Supersymmetry} \label{sect_rigid}

As mentioned in the introduction, our strategy is to start with
hypermultiplet actions with a number of abelian isometries, and to
derive the action for the scalar-tensor system by dualizing with respect
to these isometries. This method produces the most general action for
scalars coupled to tensors, where the tensors only appear through their
field strength. First, we briefly review the hypermultiplet Lagrangian,
its supersymmetry rules, and the geometry of the target space manifold.

\newpage

\textbf{Hypermultiplets}
\medskip

The Lagrangian for $n$ hypermultiplets contains $4n$ real scalars
$\phi^\A$ and $2n$ two-component spinors $\lambda^a$. Throughout this
paper, we focus on four spacetime dimensions. It is well-known
\cite{AGF,BW,DJDWKV} that for rigid $N=2$ supersymmetry, the scalar
fields $\phi^\A$ parametrize a hyperk\"ahler manifold. Such manifolds
have holonomy group contained in Sp($n$), and we denote the metric by
$G_{\A\B}(\phi)$. Our discussion closely follows \cite{DJDWKV,DWKV}, but
we change to Wess \& Bagger conventions \cite{WB}. The general form of
the Lagrangian can then be written as
 \begin{equation}
  \cL_\mathrm{H} = - \frc{1}{2}\, G_{\A\B}\, \p^\mu \phi^\A\, \p_\mu
  \phi^\B - \frc{\I}{2}\, h_{a\ab}\, \big( \la^a \si^\mu \tbs{D}_\mu
  \bla^\ab \big) + \frc{1}{4}\, W_{ab\,\ab\bb}\, \la^a \la^b\, \bla^\ab
  \bla^\bb\ ,\label{L-HM}
 \end{equation}
where $\bla^\ab=(\la^a)^*$, and $h_{a\ab}(\phi)$ and $W_{ab\,\ab\bb}
(\phi)$ are field-dependent tensors.
The covariant derivative contains a connection ${\Omega_\A}^a{}_b$,
 \begin{equation}
  D_\mu \la^a = \p_\mu \la^a + \p_\mu \phi^\A\, {\Omega_\A}^a{}_b\,
  \la^b\ ,
 \end{equation}
such that the Lagrangian is subject to two kinds of equivalence
relations, those associated with target space diffeomorphisms $\phi
\rightarrow\phi'(\phi)$, and those associated with reparametrizations
of the fermion frame $\la^a\rightarrow S^a{}_b(\phi)\,\lambda^b$, and
similarly for other quantities carrying $a,\bb$ indices. These
equivalence relations later allow us to use a convenient basis in
which the dualization procedure can be easily carried out.

The rigid supersymmetry transformations for the bosons are parametrized
by (inverse) vielbeins $\ga_{ia}^\A(\phi)$ such that
 \begin{equation}
  \de_\eps \phi^\A = \ga_{ia}^\A\, \eps^i \la^a + \bar{\ga}^i{}^\A_\ab\,
  \beps_i\, \bla^\ab\ .
 \end{equation}
Under complex conjugation, we have that $\beps_i=(\eps^i)^*$, with
$i=1,2$. The fermion transformations must be covariant with respect to
the redefinitions of the fermion frame. Hence we parametrize
 \begin{equation} \label{la_trans}
  \de_\eps \la^a + \de_\eps \phi^\A\, {\Omega_\A}^a{}_b\, \la^b =
  \I\, \p_\mu \phi^\A\, \V{\A}{ai}\, \si^\mu \beps_i\ ,
 \end{equation}
for some quantities $\V{\A}{ai}(\phi)$. The transformation on $\bla^\ab$
follows by complex conjugation.

The closure of the supersymmetry algebra and the supersymmetry
invariance of the action imposes constraints on the various quantities
which enforce the geometry of the target space to be hyperk\"ahler. This
was worked out in detail in \cite{DJDWKV,DWKV}, and we summarize it
here. First of all, the vielbeins and the tensor $h_{a\ab}$ are
covariantly constant with respect to the Levi-Civita and Sp($n$)
connections. Furthermore, one finds
 \begin{equation} \label{gammaV}
  \ga^\A_{ia}\, \V{\B}{aj} + \bar{\ga}^j{}^\A_\ab\, \bV{\B i}{\ab}
  = \de_i^j\, \de^\A_\B\ ,\qquad \ga_{ib}^\A\, \V{\A}{aj} = \de_i^j\,
  \de_b^a\ ,\qquad \bV{\A(i}{\ab}\, \ga_{j)b}^\A = 0\ .
 \end{equation}
Moreover, there is a relation between these vielbeins, the metric $G$
and the tensor $h$,
 \begin{equation} \label{G,h}
  G_{\A\B} = h_{a\bb}\, \V{\A}{ai}\, \bV{\B i}{\bb}\ ,\qquad h_{a\bb} =
  \half\, G_{\A\B}\, \ga_{ia}^\A\, \bar{\ga}^i{}_\bb^\B\ .
 \end{equation}
From this, one derives furthermore that 
 \begin{equation}\label{gV2}
  \ga^\A_{ia}\, \V{\B}{ai} = \delta^\A_\B\ .
 \end{equation}

Hyperk\"ahler manifolds have three covariantly constant complex
structures $\vec{J}$ that satisfy the quaternionic algebra. In our
notation, they are
 \begin{equation} \label{J}
  \vec{J}_{\A\B} = \I h_{b\ab}\, \bV{\A i}{\ab}\, \vec{\tau}\,{}^i{}_j
  \V{\B}{bj}\ ,
 \end{equation}
where $\vec\tau$ are the three Pauli matrices.

Finally, there are the curvature relations
\begin{equation} \label{W}
  W_{ab\,\ab\bb} = - \half\, h_{c\ab}\, \ga_{ia}^\A\, \bar{\ga}^i
  {}_\bb^\B\, R_{\A\B}{}^c{}_b = - \quart\, \ga^\A_{ia}\, \bar{\ga}^i
  {}^\B_\bb\, \ga^\C_{jb}\, \bar{\ga}^j{}^\D_\ab\, R_{\A\B\C\D}\ ,
 \end{equation}
where we have defined $R_{\A\B}=2(\p_{[\A}+\Omega_{[\A})\,\Omega_{\B]}$.
This curvature takes values in $sp(n)$ because it commutes with the
antisymmetric and covariantly constant tensor
 \begin{equation} \label{eps_ab}
  \cE_{ab} = \half \ep^{ji}\, G_{\A\B}\, \ga^\A_{ia}\, \ga^\B_{jb}\ .
 \end{equation}
Combined with the inverse $h^{\ab b}$, one can change barred indices
into unbarred ones and vice versa. One can then show that the four-fermi
tensor $W_{abcd}$ is completely symmetric in its four indices.
\bigskip

\textbf{Commuting Isometries}
\medskip

For a bosonic sigma model, one can dualize scalars into tensors if they
appear in the Lagrangian only through their derivatives. In a coordinate
invariant setting, this means that the target space has a set of abelian
Killing vectors,
 \begin{equation}
  \de_\theta \phi^\A = \theta^I k_I^\A(\phi)\ ,\qquad [ k_I\, ,\, k_J ]
  = 0\ .
 \end{equation}
Using Frobenius' theorem, one can then choose coordinates $\{\phi^I,
\phi^A\}$ such that these transformations act as constant shifts on
$\phi^I$ while leaving $\phi^A$ invariant,
 \begin{equation} \label{Frob}
  \de_\theta \phi^I = \theta^I\ ,\qquad \delta_\theta \phi^A = 0\ ,
  \qquad I = 1,\dots, n_T\ ,\qquad A = 1,\dots, 4n-n_T\ ,
 \end{equation}
and the Lagrangian depends on the former only through their field
strengths $\p_\mu\phi^I$. Vice versa, tensors can be dualized into
scalars if (but not only if) they appear only through their field
strengths, yielding commuting isometries in the corresponding sigma
model. This class of scalar-tensor models is the one we are interested
in, and within this class our dualization procedure is
general\footnote{We may assume that $G_{IJ}$ is invertible.}.

In an $N=2$ supersymmetric sigma model, dualization can only be
done if the target space isometries are triholomorphic \cite{LR}.
This means that the Killing vectors leave the complex structures
invariant, and that the isometries commute with supersymmetry.
Generically this implies that the fermions transform non-trivially,
as was worked out in detail in \cite{DWKV},
 \begin{equation}
  \de_\theta \la^a + \de_\theta \phi^\A\, {\Omega_\A}^a{}_b \la^b =
  \theta^I {t_I}^a{}_b \la^b\ ,
 \end{equation}
with matrices
 \begin{equation} \label{t-matrix}
  {t_I}^a{}_b(\phi) =\half\, \V{\B}{ai}\, \ga^\A_{ib}\, D_\A k^\B_I\ .
 \end{equation}
For Riemannian hyperk\"ahler manifolds of real dimension $4n$, the
holonomy group is contained in Sp($n$), and the group of triholomorphic
isometries must be a subgroup thereoff. The maximum number of commuting
triholomorphic isometries is therefore equal to the rank
$n$~\footnote{For pseudo-Riemannian hyperk\"ahler manifolds, the
holonomy group is generically non-compact, and there can be more than
$n$ triholomorphic abelian isometries. These cases are relevant in the
context of the superconformal tensor calculus, and examples were given
in \cite{ARV}.}. If we dualize $n_T\leq n$ scalars, we obtain a model
with $4n-n_T$ scalars coupled to $n_T$ tensors. The case of $n_T=n$
yields self-interacting $N=2$ tensor multiplets and was studied in
\cite{LR} using superspace techniques.

Using the equivalence relation based on the transformation $\la^a
\rightarrow S^a{}_b(\phi)\la^b$, one can choose an Sp($n$) frame such
that the fermions do not transform. In the basis \eqref{Frob}, this
requires
 \begin{equation}\label{int-cond}
  \p_I\, S^a{}_b = - S^a{}_c (t_I - \Omega_I)^c{}_b\ .
 \end{equation}
A solution can only be found if the integrability conditions are
satisfied. This is indeed the case, as one can check, due to the general
identity
 \begin{equation}
  D_\A {t_I}^a{}_b = k^\B_I\, {R_{\A\B}}^a{}_b\ .
 \end{equation}
Similarly, the same transformation puts the vielbeins $\ga^\A_{ia}$ and
$\V{\A}{ai}$ in a basis where they do not depend on the scalar fields
$\phi^I$, so we conclude that
 \begin{equation} \label{de_theta_la}
  \de_\theta\, \la^a = 0\ ,\qquad \de_\theta \ga^\A_{ia} = 0\ .
 \end{equation}
Through the covariant constancy of the vielbeins, $D_\B\ga^\A_{ia}=0$, 
it follows that also the Sp($n$) connection is independent of $\phi^I$, 
and, using \eqref{t-matrix}, the $I$th component is equal to
 \begin{equation}
  {\Omega_I}^a{}_b = \half\, \Gamma_{\!I\A}{}^\B\, V_\B^{ai}\,
  \ga^\A_{ib} = {t_I}^a{}_b\ .
 \end{equation}
Hence the complete hypermultiplet Lagrangian \eqref{L-HM} is invariant.
Notice that there are still residual transformations $\lambda^a
\rightarrow S^a{}_b(\phi^A)\la^b$ that define equivalence classes, since
\eqref{int-cond} is now trivially satisfied.

The conserved Noether currents of the shift symmetries \eqref{Frob},
\eqref{de_theta_la} are
 \begin{equation} \label{current}
  J^\mu_I = G_{I\A} \p^\mu \phi^\A - \I h_{b\ab}\, {\Omega_I}^b{}_a\,
  \la^a \si^\mu \bla^\ab\ ;
 \end{equation}
its divergence is given by the field equations of $\phi^I$,
 \begin{equation}
  \p_\mu J^\mu_I = \frc{\de S_\mathrm{H}}{\de \phi^I}\ .
 \end{equation}
In deriving \eqref{current}, we have used covariant constancy of
$h_{a\ab}$ and $\p_Ih_{a\ab}=0$. Combined with the covariant constancy
of \eqref{eps_ab}, this leads to the relations
 \begin{equation}
 {\Omega_I}^b{}_a h_{b\ab} + \bar{\Omega}_I{}^\bb{}_\ab h_{a\bb} = 0\
 ,\qquad {\Omega_I}^c{}_{[a}\, \cE_{b]c} = 0\ ,
 \end{equation}
which imply that the matrices ${\Omega_I}$ must be contained in
$sp(n)$.
\bigskip

\textbf{Dualization} 
\medskip

The dualization procedure can geometrically be understood as follows
\cite{HKLR,RV}: one gauges the isometries and adds Lagrange multipliers
that constrain the field strengths of the gauge potentials to be
trivial. Integrating out the multipliers sets the gauge fields to
zero\footnote{Actually, it restricts them to be pure gauge, but the
gauge parameters can be absorbed into the scalars by a field
redefinition.}, giving back the original action, whereas integrating out
the gauge potentials yields the dual action.

For an $N=2$ supersymmetric sigma model, the gauging is done by minimal
coupling to (in our case abelian) $N=2$ vector multiplets. Since these
multiplets are not propagating and just serving as a background, we can
consistently freeze the other fields of the vector multiplets (the
gauginos, scalars and auxiliary fields) to zero. We therefore only
replace the derivatives of $\phi^I$ with covariant derivatives,
 \begin{equation} \label{GCD}
  \p_\mu \phi^I \rightarrow \p_\mu \phi^I - A_\mu^I\ ,
 \end{equation}
such that to linear order the gauge fields couple to the currents
$J^\mu_I$ obtained in \eqref{current}. To maintain manifest $N=2$
supersymmetry, one has to add extra terms to the Lagrangian proportional
to the other fields of the vector multiplets. More details about these
terms are discussed below and can be found explicitly in \cite{DWKV}, or
in earlier papers on the subject. In our chosen background they vanish,
but their variations do not. For instance, the doublets of gaugino
fields $\chi^{Ii}$ vary into the field strengths according to
 \begin{equation} \label{eps-gaugino}
  \de_\eps \chi^{Ii} = \sigma^{\mu\nu} \eps^i F^I_{\mu\nu}\ .
 \end{equation}

We now introduce multipliers $B_{\mu\nu I}$ and consider the Lagrangian
 \begin{equation} \label{Lprime}
  \cL = \hat{\cL}_\mathrm{H} - H^\mu_I\, A_\mu^I\ ,
 \end{equation}
where
  \begin{equation}
  H^\mu_I = \half \ep^{\mu\nu\rho\si} \p_\nu B_{\rho\si I}
 \end{equation}
are the dual field strengths of the tensors, and where
$\hat{\cL}_\mathrm{H}$ stands for the Lagrangian \eqref{L-HM} with all
derivatives of $\phi^I$ made covariant. It is equal to the totally
gauged and supersymmetric action, where the other fields of the vector
multiplets are set to zero. $\cL$ is invariant (up to a total
derivative) under gauged isometries thanks to the Bianchi identities
of $H^\mu_I$.

Substituting the solution to the field equations for the gauge fields
$A_\mu^I$,
 \begin{equation} \label{sol_A}
  G_{IJ} A^{\mu J} = J^\mu_I - H^\mu_I\ ,
 \end{equation}
into $\cL$ yields, up to a surface term $\cL_\mathrm{surf}=-\p_\mu\phi^I
H^\mu_I$, the scalar-tensor Lagrangian
 \begin{align}
  \cL_\mathrm{T} & = \frc{1}{2}\, M^{IJ} H^\mu_I H_{\mu J} -
	\frc{1}{2}\, \cG_{AB}\, \p^\mu \phi^A\, \p_\mu \phi^B - A_A^I\,
	H^\mu_I\, \p_\mu \phi^A -\frc{\I}{2}\, h_{a\ab}\, \big( \la^a
	\si^\mu \tbs{\cD}_\mu \bla^\ab \big) \notag \\*
  & \tab + \I h_{a\ab}\, H_{\mu I} M^{IJ} {\Omega_J}^a{}_b\, \la^b
	\si^\mu \bla^\ab + \frc{1}{4}\, V_{ab\,\ab\bb}\, \la^a \la^b\,
	\bla^\ab \bla^\bb\ . \label{L_T}
 \end{align}
The tensors appearing in the kinetic terms for the bosonic fields are
 \begin{equation} \label{M}
  M^{IJ} = (G_{IJ})^{-1}\ ,\qquad \cG_{AB} = G_{AB} - G_{ AI} M^{IJ}
  G_{JB}\ ,\qquad  A_A^I = M^{IJ} G_{JA}\ .
 \end{equation}
Note that due to the isometries $\de_{\theta}\phi^I=\theta^I$, upon
dualization the hypermultiplet metric $G_{\A\B}$ splits up similarly
to a dimensional reduction on a torus:
 \begin{equation}
  G_{\A\B}\, \d\phi^\A \otimes \d\phi^\B = \cG_{AB}\, \d\phi^A \otimes
  \d\phi^B + G_{IJ}\, (\d\phi^I + A_A^I\, \d\phi^A) \otimes (\d\phi^J
  + A_B^J\, \d\phi^B)\ .
 \end{equation}
For the maximal number of commuting triholomorphic isometries, i.e.\
$n_T=n$, these metrics were studied in \cite{LR,PP}. The $A_A^I$ are
target space vector fields with abelian gauge transformations
$\de_\Lambda A_A^I=\p_A\,\Lambda^I(\phi^B)$, which leave the Lagrangian
\eqref{L_T} invariant (modulo a total derivative).

Furthermore, the covariant derivative of $\la^a$ is given by
 \begin{equation}
  \cD_\mu \la^a = \p_\mu \la^a + \p_\mu \phi^A\, \G{A}{a}{b}\, \la^b\ ,
 \end{equation}
with connection
 \begin{equation} \label{gamma}
  \G{A}{a}{b} = {\Omega_A}^a{}_b - A^I_A {\Omega_I}^a{}_b\ .
 \end{equation}
$\G{A}{a}{b}$ is invariant under the gauge transformations generated by
$\de_\Lambda$ and transforms under redefinitions of the fermion frame
like ${\Omega_A}^a{}_b$.

Finally, for the four-fermi term in the Lagrangian, we have introduced
 \begin{equation} \label{V-W}
  V_{ab\,\ab\bb} = W_{ab\,\ab\bb} + 4 h_{c\ab}\, {\Omega_I}^c{}_{(a}
  M^{IJ} {\Omega_J}^d{}_{b)}\, h_{d\bb}\ .
 \end{equation}
After conversion of the barred indices into unbarred ones by contracting
with $h^{\ab d}\cE_{dc}$, we observe that $V_{abcd}$ is still completely
symmetric.
\medskip

The supersymmetry transformations for the tensor fields leaving
$S_\mathrm{T}$ invariant can be deduced from requiring that
\eqref{Lprime} leads to an invariant action. Therefore we first need to
know how $\hat{\cL}_\mathrm{H}$ transforms. As mentioned above, this
Lagrangian is the part of the fully gauged and supersymmetric action
where the other fields of the vector multiplets are set to zero. This
action can be written as $S_\mathrm{HV}=\hat{S}_\mathrm{H}+\Delta S$,
and the only term in $\Delta S$ relevant for our considerations is the
one containing the gauginos,
 \begin{equation}
  \Delta \cL = h_{a\ab} \bar{V}_{Ii}^\ab\, \la^a \chi^{Ii} + \text{c.c.}
  + \dots\ .
 \end{equation}
The dots indicate terms proportional to the vector multiplet scalars or
auxiliary fields, whose supersymmetry variations vanish in our chosen
background. From invariance of $S_\mathrm{HV}$ and using
\eqref{eps-gaugino} we derive that (modulo a total derivative)
 \begin{equation}
  \de_\eps \hat{\cL}_\mathrm{H} = h_{a\ab} \bar{V}_{Ii}^\ab\, \eps^i
  \si^{\mu\nu} \lambda^a\, F_{\mu\nu}^I + \text{c.c.}\ .
 \end{equation}
It is now easy to write down the compensating transformation for
the tensors,
 \begin{equation}\label{susy-tensor}
  \de_\eps B_{\mu\nu I} = 2\I\, h_{a\ab} \bar{V}_{Ii}^\ab\, \eps^i
  \si_{\mu\nu} \la^a + \text{c.c.}\ .
 \end{equation}

The supersymmetry transformations of the fermions $\la^a$ in the dual
formulation are derived by making the replacement \eqref{GCD} in
\eqref{la_trans} (further corrections to the $\la^a$ transformations
due to coupling to the vector multiplets involve fields that vanish in
our background) and then substituting the solution \eqref{sol_A} for
$A_\mu^I$. This gives
 \begin{align}\label{susy-ferm}
  \de_\eps \la^a & = \I \big[ \p_\mu \phi^A\, (\V{A}{ai} - \V{I}{ai}
	M^{IJ} G_{JA}) + H_{\mu I} M^{IJ} \V{J}{ai} \big]\, \si^\mu
	\beps_i \notag \\*
  & \tab - \big[ \de_\eps \phi^\A\, {\Omega_\A}^a{}_b - 2 M^{IJ}
	\V{I}{ai} h_{c\bb}\, {\Omega_J}^c{}_b\, \beps_i \bla^\bb\,
	\big] \la^b\ .
 \end{align}
The transformations of the scalars $\phi^A$ remain the same. Notice that
this formula still contains terms proportional to $\de_\eps\phi^I$ and
therefore $\ga^I_{ia}$, which are quantities that should not appear
after dualization. Using \eqref{gammaV} and \eqref{G,h}, one can 
however express $\ga^I_{ia}$ in terms of other known quantities on the
scalar-tensor multiplet side,
 \begin{equation}
  \ga^I_{ia} = M^{IJ} h_{a\bb} \bar{V}^\bb_{Ji} - \ga^A_{ia} A^I_A\ .
 \end{equation}
By construction, the algebra of the supersymmetry transformations thus
derived closes on-shell (modulo gauge transformations of the tensors,
see below), since they are symmetries of the action $S_\mathrm{T}$.
This completes the dualization procedure.

\newpage

\textbf{Scalar-Tensor Multiplets} 
\medskip

With the insight of the previous section, we can now formulate the 
scalar-tensor multiplet Lagrangian and supersymmetry rules without
referring to hypermultiplets. The $N=2$ scalar-tensor system consists of
$n_T$ tensors $B_{\mu\nu I}$ and $4n-n_T$ scalars $\phi^A$, together
with $2n$ two-component spinors $\lambda^a$. 

The supersymmetry transformation rules are parametrized as
 \begin{align}
  \de_\eps \phi^A & = \ga_{ia}^A\, \eps^i \la^a + \bar{\ga}^i{}^A_\ab\,
	\beps_i\, \bla^\ab\ \notag \\[2pt]
  \de_\eps B_{\mu\nu I} & = 2\I\, g_{Iia}\, \eps^i \si_{\mu\nu} \la^a
	- 2\I\, \bar{g}^i_{I\ab}\, \beps_i\, \bsi_{\mu\nu} \bla^\ab
	\notag \\[2pt]
  \de_\eps \la^a & = \I \p_\mu \phi^A\, \W{A}{ai}\, \si^\mu \beps_i
	+ \I \big( H^\mu_I + k_{Ib\bb}\, \la^b \si^\mu \bla^\bb \big)
	f^{Iai}\, \si_\mu \beps_i \notag \\*
  & \tab - \de_\eps \phi^A\, \G{A}{a}{b}\, \la^b - \big( g_{Iic}\,
	\eps^i \la^c + \bar{g}_{I\cb}^i\, \beps_i \bla^\cb \big)\,
	\Gamma^{Ia}{}_b\, \la^b\ , \label{susy-dtm}
 \end{align}
for some unknown quantities $\gamma^A_{ia}$, $g_{Iia}$ etc. This is not
the most general Ansatz possible, but on-shell, and in the presence of
an action\footnote{In principle, we could relax the condition that an
action exists, in the same spirit as in \cite{Betal}. In such a setup,
it is not clear that \eqref{susy-dtm} is general enough.}, it is
sufficient since all the coefficient functions appearing here are
related by dualization to hypermultiplet quantities. Comparing with
\eqref{susy-tensor} and \eqref{susy-ferm} and using \eqref{gamma}, we
find 
 \begin{gather}
  g_{Iia} = h_{a\ab} \bar{V}_{Ii}^\ab = G_{I\A} \ga_{ia}^\A\ ,\qquad
	f^{Iai} = M^{IJ} \V{J}{ai}\ , \notag \\[2pt]
  \W{A}{ai} = \V{A}{ai} - A_A^I \V{I}{ai}\ ,\qquad \Gamma^{Ia}{}_b =
	M^{IJ} {\Omega_J}^a{}_b\ ,\qquad k_{Ib\bb} = \I h_{c\bb}\,
	{\Omega_I}^c{}_b\ . \label{dict}
 \end{gather}
If we would consider a more general Ansatz, we would find that the extra
quantities would have to vanish, or would be equivalent to the Ansatz
\eqref{susy-dtm}.

The action is parametrized by
\begin{align}
  \cL_\mathrm{T} & = \frc{1}{2}\, M^{IJ} H^\mu_I H_{\mu J} -
	\frc{1}{2}\, \cG_{AB}\, \p^\mu \phi^A\, \p_\mu \phi^B - A_A^I\,
	H^\mu_I\, \p_\mu \phi^A - \frc{\I}{2}\, h_{a\ab}\, \big( \la^a
	\si^\mu \tbs{\cD}_\mu \bla^\ab \big) \notag \\*
  & \tab + H_{\mu I} M^{IJ} k_{Ja\ab}\, \la^a \si^\mu
	\bla^\ab + \frc{1}{4}\, V_{ab\,\ab\bb}\, \la^a \la^b\, \bla^\ab
	\bla^\bb\ ,
 \end{align}
for some unknown functions $M^{IJ}$, $\cG_{AB}$, etc. We denote $H^\mu_I
=\half\ep^{\mu\nu\rho\si}\p_\nu B_{\rho\si I}$, and the covariant
derivative is given by
 \begin{equation}
  \cD_\mu \la^a = \p_\mu \la^a + \p_\mu \phi^A\, \G{A}{a}{b}\, \la^b\ .
 \end{equation}
The connection ensures covariance with respect to fermion frame
reparametrizations $\la^a\rightarrow S^a{}_b(\phi)\la^b$. 

We now require closure of the supersymmetry algebra and invariance of
the action. This imposes constraints on and relations between the
various quantities appearing in the action and supersymmetry
transformation rules, which must be equivalent with the ones that appear
on the hypermultiplet side.

For the commutator of two supersymmetries to give a translation,
we find
 \begin{gather}
  \ga_{ia}^A\, \W{A}{bj} + g_{Iia}\, f^{Ibj} = \de_i^j\, \de_a^b
	\notag \\[2pt]
  \ga_{ia}^A\, \bW{Aj}{\ab} + g_{Iia}\, \bar{f}^{I\ab}{}_j + (i
    \leftrightarrow j) = 0\ , \label{rel1}
 \end{gather}
for contractions over $A$ and $I$, and
 \begin{equation}
  \begin{pmatrix} \ga_{ia}^A\, \W{B}{aj} & \ga_{ia}^A\, f^{Jaj} \\[2pt]
  g_{Iia}\, \W{B}{aj} & g_{Iia}\, f^{Jaj} \end{pmatrix} + \text{c.c.}
  (i \leftrightarrow j) = \de_i^j \begin{pmatrix} \de_B^A & 0 \\[2pt]
  0 & \de_I^J \end{pmatrix}\ . \label{rel2}
 \end{equation}
There are further requirements coming from invariance of the action;
from the variations proportional to $\la\p^2\phi$ and $\la\p_\mu
H_{\nu}$, we find
\begin{equation}
  \cG_{AB}\, \ga_{ia}^B = h_{a\bb}\, \bW{Ai}{\bb}\ ,\qquad M^{IJ}
  g_{Jia} = h_{a\ab} \bar{f}^{I\ab}{}_i\ .
 \end{equation}
These relations imply among others that
 \begin{gather}
  \cG_{AB} = h_{a\bb}\, \W{A}{ai}\, \bW{Bi}{\bb}\ ,\qquad
    M^{IJ} = h_{a\bb}\, f^{Iai} \bar{f}^{J\bb}{}_i \notag \\[2pt]
  \de_i^j\, h_{a\bb} = \cG_{AB}\, \ga^A_{ia}\, \bar{\ga}^j{}^B_\bb
    + M^{IJ} g_{Iia}\, \bar{g}^j_{J\bb} \notag \\[2pt]
  \cE_{ab} = \half \ep^{ji}\, (\cG_{AB}\, \ga^A_{ia}\, \ga^B_{jb}\
    + M^{IJ} g_{Iia}\, g_{Jjb})\ .
 \end{gather}
Notice that as a consequence, similar to \eqref{gV2},
 \begin{equation} \label{more-contr}
  \begin{pmatrix} \ga_{ia}^A\, \W{B}{ai} & \ga_{ia}^A\, f^{Jai} \\[2pt]
  g_{Iia}\, \W{B}{ai} & g_{Iia}\, f^{Jai} \end{pmatrix} 
  = \begin{pmatrix} \de_B^A & 0 \\[2pt] 0 & \de_I^J \end{pmatrix}\ .
 \end{equation}
All the above relations can alternatively be derived by decomposing
their hypermultiplet counterparts according to the dictionary \eqref{M}
and \eqref{dict}.

Variations proportional to $\la\p_\mu\phi^A\p_\nu\phi^B$ vanish if
 \begin{equation} \label{DW}
  \cD_{\!A}\, \W{B}{ai} = -\half F_{AB}{}^I\, \bar{g}_{I\ab}^i\,
  h^{\ab a}\ .
 \end{equation}
This determines the field strength $F_{AB}{}^I=2\p_{[A}^{}A_{B]}^I$ in
terms of other quantities. The covariant derivative $\cD_{\!A}$ contains
connections $\G{A}{a}{b}$ and the Christoffel symbols $\Gamma_{\!AB}
{}^C$, built from $\cG_{AB}$. Note that the right-hand side is
antisymmetric in $A,B$, implying that $\cD^{}_{\!(A}\W{B)}{ai}=0$, and
it can be interpreted as the torsion tensor of the target space
connection.

Variations proportional to $\la H_\mu\,\p_\nu\phi$ now vanish
provided that
 \begin{align} \label{gdM}
  g_{Jia} \p_A M^{IJ} - \cG_{AB} \ga^B_{ib}\, \Gamma^{Ib}{}_a -
	h_{a\ab}\, \cD_{\!A} \bar{f}^{I\ab}{}_i & = 0 \notag \\*[2pt]
  F_{AB}{}^I \ga^B_{ia} + \cG_{AB} \ga^B_{ib}\, {\Gamma}^{Ib}{}_a -
	h_{a\ab}\, \cD_{\!A} \bar{f}^{I\ab}{}_i & = 0\ ,
 \end{align}
while variations proportional to $\la H_\mu H_\nu$ require
 \begin{equation} \label{gammadM}
  \ga^A_{ia}\, \p_A M^{IJ} = -2 M^{IK} g_{Kib} \Gamma^{Jb}{}_a\ .
 \end{equation}
Various other relations can of course be derived from this. For
instance, one may express $\Gamma^{Ia}{}_b$ in terms of other
quantities. Alternatively, one could write the field strength as
$F_{AB}{}^I=-2\,h_{b\ab}\,{\bar W}^{\ab}_{Ai}\,{\Gamma }^{Ib}{}_a\,
W_B^{ai}$.

Further consequences are used for the closure of the supersymmetry
commutator on the bosons, proportional to fermion bilinears. They can
be written as expressions for the covariant derivatives,
 \begin{equation} \label{Dg}
  \cD_{\!A} \ga^B_{ia} = -\half F_{\!AC}{}^I\, \cG^{CB} g_{Iia}\
  ,\qquad \cD_{\!A}\, g_{Iia} = -M_{IJ}\, \cG_{AB} \ga^B_{ib}\,
  \Gamma^{Jb}{}_a\ ,
 \end{equation} 
with $M_{IJ}M^{JK}=\de_I^K$. Furthermore, we find
 \begin{equation}
  k_{Ia\bb} = \I h_{c\bb}\, M_{IJ}\, \Gamma^{Jc}{}_a\ .
 \end{equation}
All these relations are consistent with the case when the tensors
decouple from the scalars. This happens when $M^{IJ}$ is constant and
the field strength $F_{AB}{}^I$ vanishes.

Using \eqref{gdM} and \eqref{Dg}, it follows that $h_{a\bb}$ and
$\cE_{ab}$ are covariantly constant with respect to the connection
$\G{A}{a}{b}$. This implies that its curvature $\cR_{AB}=2(\p_{[A}
+\Gamma_{[A})\,\Gamma_{B]}$ takes values in $sp(n)$,
 \begin{equation}
  {\cR_{AB}}^c{}_{[a}\, \cE_{b]c}=0\ ,\qquad {\cR_{AB}}^\ab{}_\bb = -
  h_{c\bb}\, h^{\ab d}\, {\cR_{AB}}^c{}_d\ .
 \end{equation}
To understand the implications of this on the holonomy of the target
space, we need to know how the Riemann curvature decomposes into its
different components. Taking a second covariant derivative of
\eqref{DW} and antisymmetrizing, we find, using \eqref{more-contr},
 \begin{equation}
  \cR_{ABCD} = h_{a\ab}\, \cR_{AB}{}^a{}_b\, \W{C}{bi}\, \bW{Di}{\ab}
  + \half M_{IJ}\, F_{C[A}{}^I F_{B]D}{}^J\ .
 \end{equation}
Hence, if $F_{AB}{}^I=0$, i.e.\ for vanishing torsion, the target space
holonomy group is restricted to be contained in Sp($n$). Note that in
general this does not imply that the target space is a hyperk\"ahler
manifold, since its dimension is $4n-n_T$. For odd $n_T$, it cannot even
admit a complex structure.

Finally, we check fermion terms of higher order in the supersymmetry 
variation of the action. Cancellation of the terms proportional to
$\la\la\bla H_\mu$ leads to 
 \begin{equation}
  V_{ab\,\ab\bb}\, \bar{f}^{I\bb}{}_i = - h_{c\ab} \big( \ga^A_{ia}\,
  \cD_{\!A}\, \Gamma^{Ic}{}_b - g_{Jia}\, [ \Gamma^I\, ,\, \Gamma^J
  ]^c{}_b - 2 g_{Jid}\, \Gamma^{Id}{}_b \Gamma^{Jc}{}_a \big)\ .
 \end{equation}
This constraint has a symmetric and antisymmetric part in $a,b$.
Vanishing of the terms proportional to $\la\la\bla\p\phi$ requires
 \begin{equation}
  V_{ab\,\ab\bb}\, \bW{Ai}{\bb} = h_{c\ab} \big( \ga^B_{ia}\, \cR_{AB}
  {}^c{}_{b} - 2 \Gamma^{Ic}{}_{a}\, \cD_{\!A}\, g_{Iib} + g_{Iia}\,
  \cD_{\!A} \Gamma^{Ic}{}_{b} \big)\ ,
 \end{equation}
which also has a symmetric and antisymmetric part. One can solve this 
for the curvature,
 \begin{equation}
  {\cR_{AB}}{}^a{}_b = h^{\ab a} V_{bc\,\ab\bb}\, \bW{Ai}{\bb} \W{B}{ci}
  - 2 h_{c\bb}\, \bW{Ai}{\bb} \W{B}{di}\, M_{IJ}\, \Gamma^{Ia}{}_d\,
  \Gamma^{Jc}{}_b\ .
 \end{equation}
Contracting these two constraints with $\bar{g}_{I\bar e}^j$ and
$\bar{\ga}^j{}^A_{\bar e}$ finally leads to an expression for the 
four-fermi tensor. Using \eqref{Dg} and \eqref{rel1}, we find
 \begin{align}
  V_{ab\,\ab\bb}\, \delta_i^j = h_{c\ab} \big( & \bar{\ga}^j{}^A_\bb\,
	\ga^B_{ia}\, \cR_{AB}{}^c{}_{b} + 2\, \delta_i^j M_{IJ}\,
	\Gamma^{Ic}{}_a \Gamma^{Jd}{}_b\, h_{d\bb} \notag \\[5pt]
  & + (g_{Iia} \bar{\ga}^j{}^A_\bb - \bar{g}^j_{I\bb} \ga^A_{ia})\,
	\cD_{\!A} \Gamma^{Ic}{}_{b} - g_{Iia}\, \bar{g}^j_{J\bb}\,
	[\, \Gamma^I ,\, \Gamma^J\, ]^c{}_b \big)\ .
 \end{align}

To end this section, we demonstrate the closure of the supersymmetry
algebra on the tensors. This is more complicated, since the commutator
only closes modulo gauge transformations and equations of motion. The
latter property is rather unusual: Recall that a supersymmetry
commutator is at most linear in derivatives, whereas bosonic equations
of motion are second-order differential equations. This is why usually
one only finds fermionic field equations, in the commutators evaluated
on the fermions. For tensors, a second derivative can be produced by
introducing an explicit spacetime coordinate dependence according to the
identity (which is valid for any vector, not just for the dual field
strength of some tensor)
 \begin{equation*}
  H_{[\mu}\, \si_{\nu]} = \si^\rho x_\rho\, \p_{[\mu} H_{\nu]} -
  \p_{[\mu} (H_{\nu]} \si^\rho x_\rho)\ ,
 \end{equation*}
as was first noticed in \cite{B1} (see also \cite{B2} for a deeper
explanation of the explicit coordinate dependence). The first term on
the right-hand side is proportional to the (linearized) field equation
of a tensor, while the second term has the form of a gauge
transformation.

Making use of this identity, we find after some algebra
 \begin{equation}
  [\, \de_\eps\, ,\, \de_{\eps'}\, ] B_{\mu\nu I} = - a^\rho \p_\rho\,
  B_{\mu\nu I} + 2\, \p_{[\mu} \Lambda_{\nu]I} + E_{IJ}\, \ep_{\mu\nu
  \rho\si}\, \frc{\de S_\mathrm{T}}{\de B_{\rho\si J}}\ ,
 \end{equation}
where
 \begin{align}
  a^\rho & = \I (\eps^i \si^\rho \beps'_i - {\eps'}^i \si^\rho
	\bar{\eps}_i ) \notag \\[2pt]
  \Lambda_{\nu I} & = a^\rho B_{\rho\nu I} + \I (\eps^i \si^\rho
	\beps'_j - {\eps'}^i \si^\rho \bar{\eps}_j)\, x_\rho\,
	\vec{\tau}\,{}^j{}_i \cdot \big[ (\vec{J}_{IA} - \vec{J}_{IJ}
	A_A^J)\, \p_\nu \phi^A \notag \\*
  & \tab + \vec{J}_{IJ} M^{JK} (H_{\nu K} + k_{Ka\ab}\, \la^a \si_\nu
	\bla^\ab) \big] \notag \\[2pt]
  E_{IJ} & = \I (\eps^i \si^\rho \beps'_j - {\eps'}^i \si^\rho
	\bar{\eps}_j)\, x_\rho\, \vec{\tau}\,{}^j{}_i \cdot
	\vec{J}_{IJ}\ .
 \end{align}
Note the explicit $x$-dependence of the gauge transformation
parameters $\Lambda_{\nu I}$ and the antisymmetric matrix $E_{IJ}$
multiplying the field equations, whose origin was explained above.
$E_{IJ}$ is field-independent (though not constant), for if the
components $\vec{J}_{IJ}=\I h^{\ab a}g_{Iia}\vec{\tau}\,{}^i{}_j
\bar{g}_{J\ab}^j$ of the complex structures \eqref{J} are independent
of the $\phi^I$, they cannot depend on the $\phi^A$ either:
 \begin{equation}
  \p_A \vec{J}_{IJ} = 3\, \p_{[A} \vec{J}_{IJ]} = 3\, D_{[A}
  \vec{J}_{IJ]} = 0\ .
 \end{equation}

\section{Local Supersymmetry} \label{sect_local}

In this section we consider hypermultiplets coupled to $N=2$
supergravity with a certain number of commuting target space isometries.
This property we then use to dualize the corresponding scalars into
tensors, just as we did for rigid supersymmetry. In this way, we find
the most general locally $N=2$ supersymmetric system of scalars and
tensors (with at most two derivatives and undeformed gauge
transformations).

For local $N=2$ supersymmetry, the $4n$ real hypermultiplet scalars
parametrize a quaternionic manifold \cite{BW}. The holonomy group of
such manifolds is contained in $\mathrm{Sp}(n)\times\mathrm{Sp}(1)$,
with a nonvanishing Sp(1) connection.

We denote the supergravity multiplet components by $e_\mu{}^m$, $A_\mu$
and $\psi_\mu^i$. The Lagrangian then reads\footnote{We define dual
field strengths $\tilde{F}^{\mu\nu}=\half\ep^{\mu\nu\rho\si}F_{\rho\si}$
and $H^\mu=\half\ep^{\mu\nu\rho\si}\p_\nu B_{\rho\si}$ as tensors, where
$\ep^{0123}=e^{-1}$.}
 \begin{align}\label{local-stm}
  e^{-1} \cL_\mathrm{H} & = - \frc{1}{2\kappa^2}\, R(e,\omega) +
	\ep^{\mu\nu\rho\si} (D_\mu \psi_\nu^i \si_\rho \bpsi_{\si i}
	+ \psi_\si^i \si_\rho D_\mu \bpsi_{\nu i}) - \frc{1}{4}\,
	\cF^{\mu\nu} \cF_{\mu\nu} \notag \\
  & \tab - \frc{\kappa}{2\2}\, (\tilde{\cF}^{\mu\nu} + \tilde{F}^{\mu
	\nu})\, (\psi_\mu^i \psi^{}_{\nu i} + \bpsi_\mu^i \bpsi^{}_{\nu
	i}) - \frc{\I}{2}\, h_{a\ab}\, (\la^a \si^\mu D_\mu \bla^\ab -
	D_\mu \la^a \si^\mu \bla^\ab) \notag \\
  & \tab - \frc{1}{2}\, G_{\A\B}\, \hat{D}^\mu \phi^\A\, \hat{D}_\mu
	\phi^\B + \kappa\, G_{\A\B} (\hat{D}_\mu \phi^\A + \p_\mu
	\phi^\A)\, (\ga^\B_{ia}\, \la^a \si^{\mu\nu} \psi_\nu^i +
	\text{c.c.}) \notag \\
  & \tab - \frc{\I\kappa}{2\2}\, \cF_{\mu\nu}\, (\cE_{ab}\, \la^a
	\si^{\mu\nu} \la^b - \text{c.c.}) - \frc{\kappa^2}{8}\,
	(\cE_{ac} \cE_{bd}\, \la^a \la^b\ \la^c \la^d + \text{c.c.})
	\notag \\
  & \tab + \frc{1}{4}\, W_{ab\,\ab\bb}\, \la^a \la^b\, \bla^\ab
	\bla^\bb\ .
 \end{align}
Here, the supercovariant field strength and supercovariant derivative
are given by
 \begin{align}
  \cF_{\mu\nu} & = F_{\mu\nu} + \2\,\I \kappa\, (\psi_\mu^i \psi^{}_{\nu
    i} - \bpsi_\mu^i \bpsi^{}_{\nu i}) \notag \\[2pt]
  \hat{D}_\mu \phi^\A & = \p_\mu \phi^\A - \kappa\, (\ga^\A_{ia}\,
    \psi_\mu^i \la^a + \bar{\ga}^i{}^\A_\ab\, \bpsi_{\mu i} \bla^\ab)\ ,
 \end{align}
while
 \begin{equation}
  D_\mu \la^a = \nabla_{\!\mu} \la^a + \p_\mu \phi^\A\, {\Omega_\A}^a
  {}_b\, \la^b\ ,\qquad D_\mu \psi_\nu^i = \nabla_{\!\mu} \psi_\nu^i +
  \p_\mu \phi^\A\, {\Omega_\A}^i{}_j\, \psi_\nu^j
 \end{equation}
are Lorentz and $\mathrm{Sp}(n)\times\mathrm{Sp}(1)$-covariant
derivatives. Notice that now the inverse vielbeins $\ga^\A_{ia}$ are
appearing explicitly in the Lagrangian. $\ga^\A_{ia}$ and $\V{\A}{ai}$
are subject to the same algebraic relations as for rigid supersymmetry,
with the difference that they are now covariantly constant with respect
to the  Levi-Civita and $\mathrm{Sp}(n)\times\mathrm{Sp}(1)$
connections. This leads to integrability relations which decompose the
Riemann curvature into an Sp($n$) and an Sp(1) part. The quaternionic
geometry relates the curvature of the the Sp(1) connection
${\Omega_\A}^i{}_j$ to the quaternionic two-forms, as it is expressed
by the relation
 \begin{equation}
  {R_{\A\B}}^i{}_j = - \ihalf \kappa^2\, \vec{J}_{\A\B} \cdot
  \vec{\tau}\,{}^i{}_j = - \kappa^2\, h_{a\ab}\, \V{[\A}{ai}\,
  \bV{\B]j}{\ab}\ .
 \end{equation}
These facts imply that the scalar fields span a negatively curved
quaternionic manifold \cite{BW}.

The four-fermi tensor $W_{ab\ab\bb}$ in \eqref{local-stm} is still
given by \eqref{W}. That there is no difference with rigid supersymmetry
can be seen from the fact that no variations of supergravity fields
contribute to the variation of the action proportional to $\la\la\bla\p
\phi$. We would like to emphasize however, that the corresponding tensor
$W_{abcd}$ is \emph{not} the same as in \cite{BW} or as the one derived
from Appendix B in \cite{Betal}. In the latter two references, one has a
completely symmetric four-index tensor. This tensor differs from ours by
a term proportional to $\cE_{a(c}\cE_{d)b}$, or, for $W_{ab\ab\bb}$, by
a term proportional to $h^{}_{a(\ab}h^t_{\bb)b}$. One can choose to
subtract this term from our $W$ and add it as a separate term in the
Lagrangian (as was done in \cite{BW}). To keep formulas shorter, we
prefer not to do so, but stress again that both approaches are
consistent and equivalent.

The action is invariant under the following local supersymmetry
transformations:
 \begin{align}
  \de_\eps e_\mu{}^m & = \I \kappa\, (\eps^i \si^m \bpsi^{}_{\mu i} -
	\psi_\mu^i \si^m \beps_i) \notag \\[2pt]
  \de_\eps A_\mu & = \2\, \I\, (\eps_i \psi_\mu^i + \beps^i \bpsi_{\mu
	i}) \notag \\[2pt]
  \de_\eps \phi^\A & = \ga_{ia}^\A\, \eps^i \la^a + \bar{\ga}^i
	{}^\A_\ab\, \beps_i\, \bla^\ab \notag \\[2pt]
  \de_\eps \la^a & = \I \hat{D}_\mu \phi^\A\, \V{\A}{ai}\, \si^\mu
	\beps_i - \de_\eps \phi^\A\, {\Omega_\A}^a{}_b\, \la^b \notag
	\\[2pt]
  \de_\eps \psi_\mu^i & = \kappa^{-1} D_\mu \eps^i + \frc{1}{2\2}\,
	\big( \cF_{\mu\nu} + \I \tilde{\cF}_{\mu\nu} + \frc{\I\kappa}
	{\2}\, \cE_{ab}\, \la^a \si_{\mu\nu} \la^b \big)\, \ep^{ij}
	\si^\nu \beps_j \notag \\*
  & \tab - \de_\eps \phi^\A\, {\Omega_\A}^i{}_j\, \psi_\mu^j\ .
 \end{align}

In order to dualize, we again need a number of commuting isometries.
Under such isometries generated by $k_I^\A$, in general not only the
$\la^a$ will transform non-trivially, but also the gravitinos, i.e.,
we have
 \begin{equation}
  \de_\theta \la^a + \de_\theta \phi^\A\, {\Omega_\A}^a{}_b \la^b =
  \theta^I {t_I}^a{}_b\, \la^b\ ,\qquad \de_\theta \psi_\mu^i +
  \de_\theta \phi^\A\, {\Omega_\A}^i{}_j \psi_\mu^j = \theta^I {t_I}^i
  {}_j\, \psi_\mu^j\ .
 \end{equation}
The two matrices ${t_I}^a{}_b$ and ${t_I}^i{}_j$ can be determined from
the requirement that the change of the vielbein $\V{\A}{ai}$ be
compensated by a combined $\mathrm{Sp}(n)\times\mathrm{Sp}(1)$
transformation,
 \begin{align}
  0 & = \cL_{k_I} \V{\A}{ai} - (t_I - k_I^\B \Omega_\B)^a{}_b\,
	\V{\A}{bi} - (t_I - k_I^\B \Omega_\B)^i{}_j\, \V{\A}{aj}
	\notag \\
  & = D_\A k_I^\B\, \V{\B}{ai} - {t_I}^a{}_b\, \V{\A}{bi} -
	{t_I}^i{}_j\, \V{\A}{aj}\ .
 \end{align}
Using the tracelessness of the $t_I$ matrices, which follows from the
relations ${t_I}^c{}_{[a}\,\cE_{b]c}={t_I}^k{}_{[i}\,\ep_{j]k}=0$, we
find
 \begin{equation}
  {t_I}^a{}_b = \frc{1}{2}\, \V{\B}{ai}\, \ga^\A_{ib}\, D_\A k^\B_I\
  ,\qquad {t_I}^i{}_j = \frc{1}{2n}\, \V{\B}{ai}\, \ga^\A_{ja}\, D_\A
  k^\B_I\ .\label{t-matrices}
 \end{equation}
The matrices ${t_I}^a{}_b$ are the same as for rigid supersymmetry, and
the matrices ${t_I}^i{}_j$ are proportional to the quaternionic moment
maps of the commuting isometries \cite{Gal}. In analogy with the rigid
case, one can now show that there is an $\mathrm{Sp}(n)\times
\mathrm{Sp}(1)$ frame such that the fermions do not transform under the
isometries and, in the Frobenius basis introduced in \eqref{Frob}, no
geometric quantities depend on $\phi^I$. In this basis, we have again
that
 \begin{equation} \label{t=omega}
  {t_I}^a{}_b = {\Omega _I}^a{}_b\ ,\qquad {t_I}^i{}_j =
  {\Omega _I}^i{}_j\ .
 \end{equation}

The conserved currents of the shift symmetries $\de_\theta\phi^I=
\theta^I$ are given by
 \begin{align}
  e^{-1} J^\mu_I & = G_{I\A} \hat{D}^\mu \phi^\A - \I h_{b\ab}\,
	{\Omega_I}^b{}_a\, \la^a \si^\mu \bla^\ab - 2 \ep^{\mu\nu\rho
	\si}\, {\Omega_I}^j{}_i\, \psi_\nu^i \si_\rho \bpsi_{\si j}
	\notag \\*
  & \tab - 2 \kappa\, (h_{a\ab} \bar{V}_{Ii}^\ab\, \la^a \si^{\mu\nu}
	\psi_\nu^i + \text{c.c.})\ .
 \end{align}
The dualization procedure now works just as in the rigid case. We first
gauge the isometries by making the replacement \eqref{GCD} in
$\cL_\mathrm{H}$ and add a Lagrange multiplier term $-e H^\mu_I A_\mu^I$
to the gauged action. The auxiliary gauge fields we then eliminate by
their equations of motion. The solution is
 \begin{equation} \label{sol_A_local}
  - G_{IJ} A^{\mu J} = H^\mu_I - e^{-1} J^\mu_I = \cH^\mu_I - G_{I\A}
  \hat{D}^\mu \phi^\A + \I h_{b\ab}\, {\Omega_I}^b{}_a\, \la^a \si^\mu
  \bla^\ab\ ,
 \end{equation}
where we have introduced the supercovariant field strengths of the
tensors:
 \begin{equation}
  \cH^\mu_I = \half \ep^{\mu\nu\rho\si} \big[ \p_\nu B_{\rho\si I} +
  4 {\Omega_I}^i{}_j\, \psi_\nu^j \si_\rho \bpsi_{\si i} - 2\I\kappa\,
  (g_{Iia}\, \psi_\nu^i \si_{\rho\si} \la^a - \text{c.c.}) \big]\ .
 \end{equation}
$\cH^\mu_I$ has to be supercovariant since all other terms in
\eqref{sol_A_local} are. This then enables us to immediately derive the
supersymmetry transformations of the tensors, without working out the
gaugino couplings and computing the compensating transformation of the
Lagrange multiplier term, as we had to do in the rigid case. If
$\de_\eps\cH^\mu_I$ is to contain only undifferentiated parameters
$\eps^i$, $\bar{\eps}_i$, then the transformation of $B_{\rho\si I}$
has to precisely cancel the inhomogeneous terms in the transformation
of the gravitinos. Thus,
 \begin{equation} \label{de_B_loc}
  \de_\eps B_{\mu\nu I} = 2\I\, g_{Iia}\, \eps^i \si_{\mu\nu} \la^a
  - 4 \kappa^{-1} {\Omega_I}^i{}_j\, \eps^j \si_{[\mu} \bpsi_{\nu]i}
  + \text{c.c.}
 \end{equation}
For the transformation of $\la^a$ in the dual formulation we find the
same expression as in the previous section, with supercovariant field
strengths in place of ordinary ones:
 \begin{align}
  \de_\eps \la^a & = \I \hat{D}_\mu \phi^A\, \W{A}{ai}\, \si^\mu \beps_i
	+ \I \big( \cH^\mu_I + k_{Ib\bb}\, \la^b \si^\mu \bla^\bb \big)
	f^{Iai}\, \si_\mu \beps_i \notag \\*
  & \tab - \de_\eps \phi^A\, \G{A}{a}{b}\, \la^b - \big( g_{Iic}\,
	\eps^i \la^c + \bar{g}_{I\cb}^i\, \beps_i \bla^\cb \big)\,
	\Gamma^{Ia}{}_b\, \la^b\ .
 \end{align}
The transformations of the bosonic fields $e_\mu{}^m$, $A_\mu$ and
$\phi^A$ do not change, so it remains to give the gravitino
transformation law, which dualization turns into
 \begin{align}
  \de_\eps \psi_\mu^i & = \kappa^{-1}\, \cD_\mu \eps^i + \frc{1}{2\2}\,
	\big( \cF_{\mu\nu} + \I \tilde{\cF}_{\mu\nu} + \frc{\I\kappa}   
	{\2}\, \cE_{ab}\, \la^a \si_{\mu\nu} \la^b \big)\, \ep^{ij}
	\si^\nu \beps_j \notag \\*
  & \tab + \kappa^{-1} \big[ \cH_{\mu I} + k_{Ia\ab}\, \la^a \si_\mu
	\bla^\ab + \kappa\, (g_{Ika}\, \psi_\mu^k \la^a + \bar{g}_{I
	\ab}^k\, \bpsi_{\mu k} \bla^\ab) \big] \Gamma^{Ii}{}_j\,
	\eps^j \notag \\*[4pt]
  & \tab - \de_\eps \phi^A\, \G{A}{i}{j}\, \psi_\mu^j - \big( g_{Ika}\,
	\eps^k \la^a + \bar{g}_{I\bb}^k\, \beps_k \bla^\bb \big)\,
	\Gamma^{Ii}{}_j\, \psi_\mu^j\ .
 \end{align}
The coefficient functions $g_{Iia}$, $\W{A}{ai}$, etc.\ appearing in the
above equations are related to hypermultiplet quantities in the same way
as in the rigid case. Hence, they satisfy the same relations
\eqref{rel1}--\eqref{more-contr}. Moreover, we have the relation
 \begin{equation}
  \Gamma^{Ii}{}_j = M^{IJ} {\Omega_J}^i{}_j
 \end{equation}
between the coefficients which appear in the supersymmetry
transformations of the gra\-vi\-tinos and tensors, respectively. 

Upon substitution of $A_\mu^I$ into the action we obtain the tensor
formulation of the theory:
 \begin{align}
  e^{-1} \cL_\mathrm{T} & = - \frc{1}{2\kappa^2}\, R - \frc{1}{4}\,
	\cF^{\mu\nu} \cF_{\mu\nu} - \frc{1}{2}\, \cG_{AB}\, \hat{D}^\mu
	\phi^A\, \hat{D}_\mu \phi^B + \frc{1}{2}\, M^{IJ} \cH^\mu_I\,
	\cH_{\mu J} - A_A^I H^\mu_I\, \p_\mu \phi^A \notag \\
  & \tab + \ep^{\mu\nu\rho\si} (\cD_\mu \psi_\nu^i \si_\rho \bpsi_{\si
	i} + \psi_\si^i \si_\rho \cD_\mu \bpsi_{\nu i}) - \frc{\I}{2}\,
	h_{a\ab}\, (\la^a \si^\mu \cD_\mu \bla^\ab - \cD_\mu \la^a
	\si^\mu \bla^\ab) \notag \\
  & \tab + \kappa\, \cG_{AB} (\hat{D}_\mu \phi^A + \p_\mu \phi^A)\,
	(\ga^B_{ia}\, \la^a \si^{\mu\nu} \psi_\nu^i + \text{c.c.}) +
	\kappa M^{IJ} \cH^\mu_I\, (g_{Jia}\, \psi_\mu^i \la^a +
	\text{c.c.}) \notag \\[2pt]
  & \tab - \frc{\kappa}{2\2}\, (\tilde{\cF}^{\mu\nu} + \tilde{F}^{\mu
	\nu})\, (\psi_\mu^i \psi^{}_{\nu i} + \bpsi_\mu^i \bpsi^{}_{\nu
	i}) - \frc{\I\kappa}{2\2}\, \cF_{\mu\nu}\, (\cE_{ab}\,\la^a
	\si^{\mu\nu} \la^b - \text{c.c.}) \notag \\
  & \tab + M^{IJ} k_{Ja\ab}\, \la^a \si^\mu \bla^\ab \big[ \cH_{\mu I}
	+ \kappa\, (g_{Iib}\, \psi_\mu^i \la^b + \text{c.c.}) \big]
	\notag \\[4pt]
  & \tab - \kappa^2 M^{IJ} (g_{Iia}\, \psi_\mu^i \la^a + \text{c.c.})\,
	(g_{Jjb}\, \la^b \si^{\mu\nu} \psi_\nu^j + \text{c.c.})
	\notag \\
  & \tab - \frc{\kappa^2}{8}\, (\cE_{ac} \cE_{bd}\, \la^a \la^b\ \la^c
	\la^d + \text{c.c.}) + \frc{1}{4}\, V_{ab\,\ab\bb}\, \la^a
	\la^b\, \bla^\ab \bla^\bb\ .
 \end{align}
The covariant derivatives $\cD_\mu$ of the fermions contain connections
$\G{A}{a}{b}$ and $\G{A}{i}{j}$, where the former is given by
\eqref{gamma} and the latter by
 \begin{equation}
  \G{A}{i}{j} = {\Omega_A}^i{}_j - A_A^I\, {\Omega_I}^i{}_j\ .
 \end{equation}
For the four-fermi terms, the tensor $V_{ab\ab\bb}$ is again determined
by the relation \eqref{V-W}, but $V_{abcd}$ is no longer completely
symmetric. Note also that the $M^{IJ}\cH^\mu_I\,\cH_{\mu J}$ term gives
rise to a similar correction to the four-gravitino coupling of the
hypermultiplet action.

Finally, let us derive a relation for the target space curvature in the
tensor formulation. The covariant derivatives of the vielbeins $\W{A}
{ai}$, etc.\ given in \eqref{DW} and \eqref{Dg} for rigid supersymmetry
now include in addition Sp(1) connection coefficients. In particular,
$\cD^{}_{\!A}\W{B}{ai}$ is still given by the right-hand side of
\eqref{DW}, but we now have
 \begin{align}
  \ga^A_{ia}\, \p_A M^{IJ} = -2 M^{IK} (g_{Kib} \Gamma^{Jb}{}_a +
	g_{Kja} \Gamma^{Jj}{}_i)\ \notag \\[2pt]
  \cD_{\!A}\, g_{Iia} = -M_{IJ}\, \cG_{AB}\, (\ga^B_{ib}\, \Gamma^{Jb}
	{}_a + \ga^B_{ja}\, \Gamma^{Jj}{}_i)\ .
 \end{align}
These relations further constrain $\Gamma^{Ia}{}_b$ and $\Gamma^{Ii}{}_j$, 
in a way consistent with \eqref{t-matrices} and \eqref{t=omega}.

By computing the commutator of two covariant derivatives on $W_A^{ai}$, we 
derive the following curvature relation,
 \begin{equation}
  \cR_{ABCD} = h_{a\ab}\, \cR_{AB}{}^a{}_b\, \W{C}{bi}\, \bW{Di}{\ab} +
  \cR_{AB}{}^i{}_j\, \W{C}{aj} h_{a\ab} \bar{W}_{\!Di}^\ab + \half
  M_{IJ}\, F_{C[A}{}^I F_{B]D}{}^J\ .
 \end{equation}
Due to covariant constancy of $\cE_{ab}$ and $\ep_{ij}$ with respect to
the connections $\G{A}{a}{b}$ and $\G{A}{i}{j}$, the corresponding
curvatures take values in $sp(n)$ and $sp(1)$, respectively. We conclude
that if $F_{AB}{}^I$ vanishes the holonomy group of the target space is
contained in $\mathrm{Sp}(n)\times\mathrm{Sp}(1)$. In the next section
we present an example with vanishing $F_{AB}{}^I$.

\section{The Universal Hypermultiplet and its Dual Double-Tensor
Multiplet} \label{sect_DTM}

In order to illustrate the results of the previous section, let us now
apply the dualization to the example of the universal hypermultiplet
coupled to $N=2$ supergravity. This multiplet arises in Calabi-Yau
threefold compactifications of Type II supergravities, and contains four
real scalar fields that parametrize the homogeneous quaternion-K\"ahler
target space SU(1,\,2)/U(2) \cite{CFG}. Out of the isometry group 
SU(1,\,2), we may pick two commuting U(1) isometries that may be used to 
dualize one or two scalars into tensors. In fact, in the case of Type IIA 
the universal hypermultiplet arises after dualization of a tensor multiplet, 
while for Type IIB it follows from a double-tensor multiplet.

In this section we shall reverse the latter dualization and derive the
double-tensor multiplet by dualizing the pseudoscalars (i.e., from a IIB
perspective, the axion and one of the RR scalars) in the universal
hypermultiplet, according to the procedure of section \ref{sect_local}.
In this way, we obtain the fermionic terms in the action and the
supersymmetry rules for the double-tensor multiplet, which were
previously unknown. We should mention, however, that in the framework of
compactified Type IIB supergravity the double-tensor multiplet is
accompanied by $h_{1,1}>0$ tensor multiplets, and it is inconsistent to
truncate them. The case of the pure double-tensor multiplet, as
presented here, should be understood as the mirror theory of Type IIA
supergravity on a rigid ($h_{1,2}=0$) Calabi-Yau, where no tensor
multiplets occur.

We parametrize the target space of the universal hypermultiplet by four
real scalars $\phi^\A=(\phi,\,\chi,\varphi,\si)$. Its metric is given in
terms of the matrix-valued vierbein 1-form\footnote{In this section we
choose constant tensors $h_{a\ab}=\de_{a\ab}$, $\ep_{12}=\cE_{12}=-1$.
Furthermore, we set $\kappa^{-1}=\2$ and rescale the transformation
parameters $\eps^i\rightarrow\2\eps^i$.}
 \begin{equation}
  \d\phi^\A\, \V{\A}{ai} = \frc{1}{\2} \begin{pmatrix} \e^{-\phi/2}
  (\d\chi - \I\, \d\varphi) & \d\phi + \I\, \e^{-\phi} (\d\si + \chi
  \d\varphi) \\[2pt] - \d\phi + \I\, \e^{-\phi} (\d\si + \chi \d\varphi)
  & \e^{-\phi/2} (\d\chi + \I\, \d\varphi) \end{pmatrix}\ ,
 \end{equation}
as $G=\mathrm{tr}\,(V\otimes V^\dag)$. The bosonic part of the action
then reads explicitly
 \begin{align}
  e^{-1} \cL_\mathrm{UH} & = - R - \frc{1}{4}\, F^{\mu\nu} F_{\mu\nu}
	- \frc{1}{2}\, \p^\mu \phi\, \p_\mu \phi - \frc{1}{2}\,
	\e^{-\phi} \big( \p^\mu \chi\, \p_\mu \chi + \p^\mu \varphi\,
	\p_\mu \varphi \big) \notag \\*
 & \tab - \frc{1}{2}\, \e^{-2\phi} \big( \p_\mu \si + \chi \p_\mu
	\varphi \big)^2 + \dots \ .
 \end{align}
The global SU(1,\,2) isometry group has an obvious $\mathrm{U}(1)\times
\mathrm{U}(1)$ subgroup, which is generated by constant shifts of
$\phi^I=(\varphi,\,\si)$ and which does not act on the other scalars
$\phi^A=(\phi,\chi)$. With our parametrization of the vierbein, no
quantity will depend on the $\phi^I$, so all requirements for the
dualization are fulfilled.

Since we consider $n=1$ hypermultiplets, the holonomy group has two
Sp(1) factors. Their connections follow from covariant constancy of the
vierbein (cf.\ \cite{S}):
 \begin{align}
  \d\phi^\A\, {\Omega_\A}^a{}_b & = \frc{3\I}{4}\, \e^{-\phi}
	\begin{pmatrix} -(\d\si + \chi \d\varphi) & 0 \\[2pt] 0 &
	(\d\si + \chi \d\varphi) \end{pmatrix}\ , \notag \\[4pt]
  \d\phi^\A\, {\Omega_\A}^i{}_j & = \frc{\I}{4}\, \e^{-\phi}
	\begin{pmatrix} (\d\si + \chi \d\varphi) & 2\, \e^{\phi/2}
	(\d\varphi + \I\, \d\chi) \\[2pt] 2\, \e^{\phi/2} (\d\varphi
	-\I\, \d\chi) & -(\d\si + \chi \d\varphi) \end{pmatrix}\ .
 \end{align}
Using \eqref{W}, which is valid also in the local case, the four-fermi
term with coefficient $W_{ab\,\ab\bb}$ can be expressed in terms of
the symmetric matrix $\Sigma^{ab}=\la^a\la^b$ as
 \begin{align}
  \frc{1}{4}\, W_{ab\,\ab\bb}\, \la^a \la^b\, \bla^\ab \bla^\bb & =
	-\frc{1}{8}\, \mathrm{tr}\, \big[ \ga^\A \Sigma\, R_{\A\B}^t\,
	h\, (\ga^\B \Sigma)^\dag \big] = \frc{3}{16}\, \mathrm{tr}\,
	\big[ \tau^3 \Sigma \tau^3 \bar{\Sigma} \big] \notag \\[2pt]
  & = \frc{3}{16}\, \big( \la^1 \la^1\, \bla^1 \bla^1 - 2\, \la^1
	\la^2\, \bla^1 \bla^2 + \la^2 \la^2\, \bla^2 \bla^2 \big)\ .
 \end{align}

We now dualize the scalars $\phi^I$ into two tensors $B_{\mu\nu I}$
along the lines in the previous sections. The metrics of the
double-tensor multiplet follow from \eqref{M} and read
 \begin{equation}
  M^{IJ} = \e^{\phi} \begin{pmatrix} 1 & -\chi \\[2pt] -\chi & \e^\phi
  + \chi^2 \end{pmatrix}\ ,\qquad \cG_{AB} = \begin{pmatrix} 1 & 0
  \\[2pt] 0 & \e^{-\phi} \end{pmatrix}\ ,\qquad  A_A^I = 0\ .
 \end{equation}
The last equality is a consequence of $G_{\A\B}$ being block-diagonal,
$G_{AI}=0$, which simplifies the dualization considerably. For the
scalar zweibeins in the dual formulation we find
 \begin{equation}
  \gamma^\phi_{ia} = (W_\phi^{ai})^\dag = \frc{1}{\2} \begin{pmatrix}
  0 & -1 \\[2pt] 1 & 0 \end{pmatrix}\ ,\qquad \gamma^\chi_{ia} = \e^\phi
  (W_\chi^{ai})^\dag = \frc{1}{\2}\, \e^{\phi/2} \begin{pmatrix} 1 & 0
  \\[2pt] 0 & 1 \end{pmatrix}\ ,
 \end{equation}
while the tensor zweibeins are given by
 \begin{equation}
  g_{1\,ia} = -\frc{\I}{\2}\, \e^{-\phi} \begin{pmatrix} -\e^{\phi/2} &
  \chi \\[2pt] \chi & \e^{\phi/2} \end{pmatrix}\ ,\qquad g_{2\,ia} = -
  \frc{\I}{\2}\, \e^{-\phi} \begin{pmatrix} 0 & 1 \\[2pt] 1 & 0
  \end{pmatrix}\ ,
 \end{equation}
(where $I=1$ refers to $\varphi$ and $I=2$ to $\si$) and
 \begin{equation}
  f^{1\,ai} = \frc{\I}{\2}\, \e^{\phi/2} \begin{pmatrix} -1 & 0 \\[2pt]
  0 & 1 \end{pmatrix}\ ,\qquad f^{2\,ai} = \frc{\I}{\2}\, \e^{\phi/2}
  \begin{pmatrix} \chi & \e^{\phi/2} \\[2pt] \e^{\phi/2} & -\chi
  \end{pmatrix}\ .
 \end{equation}
One may check that these quantities satisfy the relations
\eqref{rel1}--\eqref{more-contr}. In particular, since $\ga^\chi_{ia}$
and $W_\chi^{ai}$ are both real and proportional to the unit matrix,
the off-diagonal terms in \eqref{rel2} and \eqref{more-contr} imply
that both $g_{Iia}$ and $f^{Iai}$ must be antihermitean and traceless,
which is clearly the case.

The target space connections for the double-tensor multiplet are
particularly simple:
 \begin{equation}
  \G{A}{a}{b} = 0\ ,\qquad \G{\phi}{i}{j} = 0\ ,\qquad \G{\chi}{i}{j}
  = \frc{1}{2}\, \e^{-\phi/2} \begin{pmatrix} 0 & -1 \\[2pt] 1 & 0
  \end{pmatrix}\ .
 \end{equation}
Since $A_A^I=0$, the scalar zweibeins $\W{A}{ai}$, $\ga^A_{ia}$ are
covariantly constant with respect to these connections. The other
quantities that derive from the connections of the hypermultiplet are
the gra\-vitino coefficients in the supersymmetry transformations of
the tensors \eqref{de_B_loc}
 \begin{gather}
  {\Omega_1}^i{}_j = \frc{\I}{4}\, \e^{-\phi} \begin{pmatrix} \chi &
	2\, \e^{\phi/2} \\[2pt] 2\, \e^{\phi/2} & -\chi \end{pmatrix}\
	,\qquad {\Omega_2}^i{}_j = \frc{\I}{4}\, \e^{-\phi}
	\begin{pmatrix} 1 & 0 \\[2pt] 0 & -1 \end{pmatrix}\ ,
 \end{gather}
and the coefficients in the transformations of the fermions
 \begin{gather}
  \Gamma^{1\,a}{}_b = 0\ ,\qquad \Gamma^{2\,a}{}_b = -\frc{3\I}{4}\,
	\e^\phi \begin{pmatrix} 1 & 0 \\[2pt] 0 & -1 \end{pmatrix}\ ,
	\notag \\[4pt]
  \Gamma^{1\,i}{}_j = \frc{\I}{2}\, \e^{\phi/2} \begin{pmatrix} 0 & 1
	\\[2pt] 1 & 0 \end{pmatrix}\ ,\qquad \Gamma^{2\,i}{}_j = -
	\frc{\I}{4}\, \e^{\phi/2} \begin{pmatrix} -\e^{\phi/2} & 2
	\chi \\[2pt] 2 \chi & \e^{\phi/2} \end{pmatrix}\ .
 \end{gather}

Just as in the universal hypermultiplet, the four-$\la$ terms come with
field-independent coefficients,
 \begin{equation} \label{cR_DTM}
  \frc{1}{4}\, V_{ab\,\ab\bb}\, \la^a \la^b\, \bla^\ab \bla^\bb = -
  \frc{3}{8}\, \big( \la^1 \la^1\, \bla^1 \bla^1 - 2\, \la^1 \la^2\,
  \bla^1 \bla^2 + \la^2 \la^2\, \bla^2 \bla^2 \big)\ .
 \end{equation}
We now have determined all ingredients needed to write down the complete
action and supersymmetry transformations for the double-tensor multiplet
coupled to supergravity. We refrain from actually doing so, however,
since for applications of this result it is advantageous to keep the
vielbeins and connection coefficients in matrix form as given above and
not to explicitly perform the sums over the Sp($n$) and Sp(1) indices.
Let us just display the linearized supersymmetry transformations of the
fermions, as these are the most relevant ones in studying BPS solutions
to the theory:
 \begin{equation}
  \de_\eps \begin{pmatrix} \la^1 \\[2pt] \la^2 \end{pmatrix} =
  \begin{pmatrix} \I \e^{-\phi/2} \p_\mu \chi + \e^{\phi/2}
  \hat{H}_{\mu1} & \I \p_\nu \phi - \e^\phi H_{\nu2} \\[2pt] -\I
  \p_\mu \phi - \e^\phi H_{\mu2} & \I	\e^{-\phi/2} \p_\nu \chi -
  \e^{\phi/2} \hat{H}_{\nu1} \end{pmatrix} \begin{pmatrix} \si^\mu
  \bar{\eps}_1 \\[2pt] \si^\nu \bar{\eps}_2 \end{pmatrix} + \dots \ ,
 \end{equation}
for the hyperinos, and
 \begin{align}
  \de_\eps \begin{pmatrix} \psi_\mu^1 \\[2pt] \psi_\mu^2 \end{pmatrix}
	& = \begin{pmatrix} 2 \nabla_{\!\mu} + \ihalf \e^\phi H_{\mu2} &
	- \e^{-\phi/2} \p_\mu \chi + \I \e^{\phi/2} \hat{H}_{\mu1}
	\\[2pt] \e^{-\phi/2} \p_\mu \chi + \I \e^{\phi/2} \hat{H}_{\mu1}
	& 2 \nabla_{\!\mu} - \ihalf \e^\phi H_{\mu2} \end{pmatrix}
	\begin{pmatrix} \eps^1 \\[2pt] \eps^2 \end{pmatrix} \notag \\
  & \tab + \frc{1}{2}\, (F_{\mu\nu} + \I \tilde{F}_{\mu\nu})
	\begin{pmatrix} \si^\nu \bar{\eps}_2 \\[2pt] -\si^\nu
	\bar{\eps}_1 \end{pmatrix} + \dots \ ,
 \end{align}
for the gravitinos.

Note that $H^\mu_1$ always appears in the combination $\hat{H}^\mu_1
\equiv H^\mu_1-\chi H^\mu_2$ in the transformations. This is due to a
global symmetry of the double-tensor multiplet, which acts on the
scalars by a constant shift of $\chi$ and on the tensors such that
$\hat{H}^\mu_1$ and $H^\mu_2$ are invariant \cite{TV} (in fact, this
applies to the full supercovariant field strengths $\cH^\mu_I$). There
is another global symmetry, which acts on $\tau\equiv\chi+2\I\,\e^{\phi
/2}$ by $\tau\rightarrow\e^\alpha\tau$ and on the tensors by $B_{\mu\nu
I}\rightarrow\e^{-I\alpha}B_{\mu\nu I}$ (no sum). These symmetries do
not act on the $\la^a$ thanks to our choosing $h_{a\ab}$ and $\cE_{ab}$
constant, and they explain why the four-fermi term \eqref{cR_DTM} is
independent of the scalars.
\bigskip

\textbf{Acknowledgments} 
\medskip

UT is supported by FWF Grant P15553-N08 and was supported by DFG Grant
TH~786/1-2  during the initial stage of this work.

\raggedright

\end{document}